\documentclass[aps,showpacs,superscriptaddress,nofootinbib,amsmath,amssymb]{revtex4}
 \usepackage{graphicx}
\usepackage{dcolumn}
\usepackage{bm}
\usepackage{lipsum}
\usepackage{adjustbox}
\usepackage{multirow}
\usepackage{amsmath}
\usepackage{amssymb}
\usepackage{gensymb}
\usepackage{textcomp}
\usepackage{enumerate}

\graphicspath{./figs}
\begin{document}
\title{Nuclear effects on tau lepton polarization in charged current deep inelastic $\nu_\tau/\bar\nu_\tau-A$ scattering}
\author{F. Zaidi}
\author{M. Sajjad Athar}
\author{S. K. Singh}
\affiliation{Department of Physics, Aligarh Muslim University, Aligarh - 202002, India}
\begin{abstract}
We have studied the tau-lepton polarization in the charged current $\nu_\tau/\bar\nu_\tau$ induced deep inelastic scattering (DIS) from the free nucleon as well as off 
the nuclear targets that are being used in ongoing and 
proposed experiments such as IceCube, DUNE, etc. 
For the free nucleon target, the differential scattering cross sections are obtained by taking into account the non-perturbative effect like target mass corrections (TMC) and the perturbative effect
like the evolution of the parton densities at the next-to-leading order (NLO) in the four flavor $\overline{\textrm{MS}}-$scheme. In the case of nucleons bound inside a nuclear target, we have incorporated the nuclear medium effects
such as Fermi motion, binding energy and nucleon correlations, through the use of nucleon spectral function. 
 We shall present the results for the differential scattering cross sections and the longitudinal and transverse components of the tau-lepton polarization assuming time reversal invariance.
 
\end{abstract}
\maketitle

\section{Introduction}
In the last two decades, the particle physics community has put in appreciable efforts to study the third generation of the lepton family
i.e. the charged lepton $\tau^+(\tau^-)$ and its neutral partner $\bar\nu_\tau(\nu_\tau)$~\cite{MammenAbraham:2022xoc, Pich:2020qna}. Though, the $\tau-$lepton was discovered 
for the first time in 1975 by the SLAC-LBL collaboration~\cite{Perl:1975bf},
it took almost 25 years thereafter to detect $\nu_\tau$. The $\nu_\tau$ was detected by the DONUT collaboration~\cite{DONUT:2000fbd} in 2000 based on the observation of the four $\nu_\tau$ events 
in the charged current interaction channels. Later on, the OPERA experiment at CERN~\cite{OPERA:2015wbl} observed $\nu_\tau$ events in the direct appearance mode in the $\nu_\mu \to \nu_\tau$ oscillation channel. 
The observation of charged current $\nu_\tau$ interactions is more difficult as compared to the $\nu_e$ and $\nu_\mu$ interactions due to 
the short life of $\tau-$lepton ($2.9 \times 10^{-13}$ seconds) produced in the final state. Till date, the $\tau-$lepton events with limited statistics have been observed 
in accelerator based experiments such as DONUT (9 $\nu_\tau$ events)~\cite{DONUT:2000fbd, DONuT:2007bsg}, OPERA (10 $\nu_\tau$ events)~\cite{OPERA:2015wbl, OPERA:2018nar} 
and NOMAD (9 $\nu_\tau$ events)~\cite{NOMAD:2001xxt}, while 291 $\nu_\tau$ candidates are reported by the atmospheric neutrino based SuperK experiment~\cite{Super-Kamiokande:2017edb}, and 
1806 candidates from the astrophysical sources are identified in the IceCube experiment~\cite{IceCube:2019dqi, IceCube:2020fpi}. So far, all the available experimental measurements have large statistical and systematic errors 
of about 30\%-50\% which are mainly because of the low statistics and the experimental uncertainties~\cite{DsTau:2019wjb}. It has been planned to study the $\tau-$lepton production with improved statistics via the
decay of $D_s-$meson ($D_s \to \tau\;\nu_\tau$) in the DsTau~\cite{DsTau:2019wjb} and SHiP~\cite{SHiP:2015vad} experiments, by using emulsion detector
in FASER$\nu$~\cite{FASER:2019dxq, FASER:2020gpr, Jodlowski:2020vhr, FASER:2023zcr} and SND@LHC~\cite{SHiP:2020sos, SNDLHC:2022ihg} experiments
as well as through the $\nu_\mu (\bar\nu_\mu)\to \nu_\tau (\bar\nu_\tau)$ oscillation 
channel in the DUNE~\cite{DeGouvea:2019kea, DUNE:2020ypp}, IceCube upgrade~\cite{Ishihara:2019aao, IceCube-Gen2:2020qha} and T2HK~\cite{Hyper-Kamiokande:2018ofw, Yu:2018atd} experiments that cover
a wide energy spectrum of neutrino. DUNE is the 
only accelerator based experiment which has ability to collect and accurately reconstruct the $\nu_\tau$ charged current events. It is expected that at the DUNE, $\nu_\tau$ events in the
appearance mode would be 
between 100 to 1000~\cite{DeGouvea:2019kea, DUNE:2020ypp}. An important goal of the DUNE and HyperK experiments is to determine the neutrino mass hierarchy (normal or inverted) which rely on the $\nu_e$ flux measurements, however, it would have
uncertainty arising due to $\nu_\mu(\bar\nu_\mu) \rightleftharpoons \nu_\tau (\bar\nu_\tau)$ oscillation giving rise to tauons which decay to electrons (i.e. $\tau \rightarrow e$ decays).
The study of $\tau-$lepton and its neutral partner $\nu_\tau$ is important as it is a compelling probe to test the lepton flavor universality (LFU) which 
is one of the principal assumptions of the standard 
model of particle physics~\cite{Weinberg:1967tq, Salam:1968rm, Glashow:1970gm}. However, this assumption has been questioned by the observation of asymmetry measurements in the decays of
$B$-mesons into the semi-leptonic final states (e.g. $B^0 \rightarrow D^{\ast -}\tau^+ \nu_\tau\;;\;\;\tau^+ \rightarrow \pi^+\pi^-\pi^+(\pi^0)\bar\nu_\tau$) at the
LHCb experiment~\cite{LHCb:2014vgu, LHCb:2015gmp, LHCb:2023cjr}. Earlier, the $B-$mesons decay into the $\tau-$lepton has been investigated by the dedicated 
BABER~\cite{BaBar:2007hvx, BaBar:2012obs} and Belle~\cite{Belle:2007qnm, Belle:2009xqm, Belle:2010tvu} experiments. The search for lepton flavor violation may open a window to discover new physics. 
The upgraded Belle experiment (Belle-II), a next generation flavor factory,
at the SuperKEK~\cite{Belle-II:2018jsg} also aims to study the $\tau-$lepton decay modes which are relevant to test the standard model assumptions
as well as to explore the new physics beyond the SM. It is important to mention that among the charged leptons ($e^\pm$, $\mu^\pm$, $\tau^\pm$), $\tau-$lepton is the only one which 
decays into hadrons and thus may prove to be a unique source in the study of both inclusive as well as exclusive processes corresponding to the 
perturbative and non-perturbative energy regimes, respectively~\cite{Portoles:2022mxb}. Furthermore, a thorough understanding of the $\tau-$lepton properties is required for the accurate measurement 
of the neutrino oscillation parameters, to determine the neutrino mass hierarchy, and to reduce the existing uncertainty in the neutrino-nucleon and neutrino-nucleus scattering cross
section measurements~\cite{Lipari:1994pz, Conrad:2010mh, Ansari:2020xne, Zaidi:2021iam}, as well as 
in the $\nu_\tau$ observation from astrophysical sources~\cite{IceCube:2015vkp, VanEijk:2018wps, Song:2020nfh}.  

Due to its heavy mass ($m_\tau=1.78$ GeV), $\tau-$lepton produced in the charged current $\nu_\tau$ interactions through various reaction channels like the quasielastic, the inelastic, and the
deep inelastic scattering, are not fully polarized in the energy region of a few GeV ($E_{\nu_\tau} \lesssim 10$ GeV). It is difficult to study
the reaction $\nu_\tau (\bar\nu_\tau) +_{Z}^{A}X \rightarrow \tau^- (\tau^+) + Y$ via direct detection mode due to its short life time, and $\tau-$lepton is identified by the 
observation of its decay products such as leptons and 
pions whose decay rates and topologies depend upon the production cross section and the polarization of the $\tau-$leptons. 
The effect of $\tau-$lepton polarization should be studied on the $\tau-$lepton production signal, and for 
estimating the background events in $\nu_\mu \rightarrow \nu_e$ appearance channel, etc.
Since the oscillation amplitude of $\nu_\mu \to\nu_\tau$ is larger than $\nu_\mu \to\nu_e$ and the branching ratio of $\tau^\pm \rightarrow e^\pm\;X$
is relatively large, the electron production through $\nu_\mu \rightarrow \nu_\tau \rightarrow \tau \rightarrow e$ reaction can be significant and would contribute to the background events.
It is important to point out that 
the present neutrino event generators such as GENIE~\cite{GENIE:2021zuu, Andreopoulos:2015wxa}, GiBBU~\cite{Leitner:2006ww, Buss:2011mx},
NuWro~\cite{Golan:2012rfa}, NEUT~\cite{Hayato:2021heg} assume the final state $\tau-$lepton to be purely left-handed and simulate $\nu_\tau$ events just like in 
the case of $\nu_e$ and $\nu_\mu$ events, which is not appropriate to determine the genuine number of $\nu_\tau$ events in the energy region of a few GeV~\cite{Isaacson:2023gwp}. 
In the literature, various studies are available for the $\tau-$lepton polarization in $\nu_\tau$-nucleon charged current interaction through different reaction channels such as
the quasielastic, the inelastic and the deep inelastic scattering processes. For example, in the case of CC $\nu_\tau-N$ quasielastic scattering process, the polarization of the
final state $\tau-$lepton has been studied by various authors~\cite{Hagiwara:2003di, Hagiwara:2004gs, Aoki:2005wb, Kuzmin:2003ji, Kuzmin:2004ke, Kuzmin:2004yb, Kurek:2005uz, 
Fatima:2020pvv, Fatima:2022tlf, SajjadAthar:2022pjt} while for the inelastic processes it is discussed by some of these authors in 
Refs.~\cite{Hagiwara:2003di, Kuzmin:2003ji, Hagiwara:2004gs, Aoki:2005wb, Kurek:2005uz, Graczyk:2017rti}.

The $\tau-$lepton polarization for the deep inelastic $\nu_\tau(\bar\nu_\tau)$ scattering off the free nucleon target has been discussed 
in Refs.~\cite{Albright:1974ts, Hagiwara:2003di, Kuzmin:2003ji, Hagiwara:2004gs, Graczyk:2004vg, Bourrely:2004iy, Aoki:2005wb}. However, the results reported for 
the $\tau-$lepton production cross sections and its polarization observables suffer from some uncertainties arising due
to: {\bf (i)} the adopted choice of different limits for the kinematic constrains on $W$ and $Q^2$ to demarcate the onset of DIS, and {\bf (ii)} the use of different parameterizations
of the nucleon structure functions. For example,
in Refs.~\cite{Hagiwara:2003di, Hagiwara:2004gs}, authors have evaluated the DIS cross sections and the degree of polarization at NLO  by directly
using the PDFs grids of MRST parameterization~\cite{Martin:2001es}, i.e.,
without considering $Q^2$ evolution of the parton densities from the leading order (LO) to next-to-leading order (NLO). These authors also put constraints on the four momentum transfer square such that 
the nucleon structure functions are evaluated at a fixed value of $Q^2=Q_0^2=1.25$ GeV$^2$ for all $Q^2<Q_0^2$. While Graczyk~\cite{Graczyk:2004vg} uses GRV98 PDFs parameterization~\cite{Gluck:1998xa}
and freezes $Q^2=Q_0^2=0.8$ GeV$^2$, without and with the Bodek-Yang corrections~\cite{Bodek:2002vp} to study the cross sections and polarization of the produced $\tau-$lepton in the $\nu_\tau-N$ scattering.
Moreover, in many studies, different values of the kinematical cut on the center of mass (CoM) energy ($W$) to define the onset of the DIS process are taken into 
consideration, for example, $W\ge 1.4$ GeV is taken by 
Hagiwara et al.~\cite{Hagiwara:2003di} and $W\ge M+m_\pi$ is taken by Graczyk~\cite{Graczyk:2004vg}, while the most widely used neutrino Monte Carlo event generators like 
GENIE~\cite{Andreopoulos:2015wxa}
and NEUT~\cite{Hayato:2021heg} take $W\ge 1.7$ GeV and $W\ge 2$ GeV, respectively. In the MINERvA experiment at the Fermilab~\cite{MINERvA:2016oql}, which covers a wide energy spectrum and 
performs EMC~\cite{EuropeanMuon:1983wih} kind of measurements to study the nuclear medium modifications in (anti)neutrino induced DIS processes off different 
nuclear targets, the kinematic region of $W\ge 2$ GeV and $Q^2\ge 1$ GeV$^2$ is considered to be the region of 
true DIS events~\cite{MINERvA:2014rdw, MINERvA:2016oql}. 

Since most of the neutrino experiments are performed using medium or heavy nuclear targets, a better understanding of 
the nuclear medium effects is crucial specially in the low and intermediate energy regions. Therefore, it is also important to 
study the impact of the nuclear medium effects on the polarization 
observables of the final state $\tau-$lepton which may provide a complementary way to test the nuclear models available for the scattering cross sections.
In the literature, the nuclear medium effects in the $\tau-$lepton polarization for the quasielastic scattering process are discussed in various studies by Graczyk~\cite{Graczyk:2004uy},
Valverde et al.~\cite{Valverde:2006yi}, Lagoda et al.~\cite{Lagoda:2007zz}, Amaro et al.~\cite{Amaro:2009ka}, Sobczyk et al.~\cite{Sobczyk:2019urm},
Hernandez et al.~\cite{Hernandez:2022nmp}, Isaacson et al.~\cite{Isaacson:2023gwp}, etc., 
while there is hardly any work, where $\tau-$lepton polarization effects have been analyzed in $\nu_\tau(\bar\nu_\tau)$ induced inelastic and deep inelastic scattering processes 
on nuclear targets with nuclear medium effects.

In this work, for the first time, the effects of the nuclear medium have been explicitly taken into account in the study of the $\tau-$lepton polarization in the charged 
current induced $\nu_\tau(\bar\nu_\tau)$ deep inelastic scattering
on nuclear targets. Since the energy and angular distributions of the $\tau-$lepton decay products are sensitive to the polarization of the produced $\tau-$particles, we have studied in detail
the non-vanishing components of the $\tau-$lepton polarization and the effect of the nuclear medium on these components. Assuming the time reversal invariance, only the longitudinal ($P_L$) and 
transverse ($P_T$) components of polarization are non-vanishing, and the component perpendicular to the reaction plane $(P_{P})$ vanishes. The nuclear medium effects on the polarization components
$P_L$ and $P_T$ have been studied using the methods applied earlier to study the nuclear medium effects in the calculations of differential and total cross sections in the DIS processes induced by
$\nu_\mu(\bar\nu_\mu)$ and $\nu_\tau(\bar\nu_\tau)$ on various nuclei~\cite{Zaidi:2019asc, Zaidi:2021iam, Ansari:2021cao, AtharSajjad:2022ipr, Zaidi:2022qdp}. All the numerical calculations for scattering cross sections and polarization components 
are performed by considering the kinematic limits of $W\ge 2$ GeV and $Q^2\ge 1$ GeV$^2$, which at present are generally taken to be the limits of the true DIS region~\cite{SajjadAthar:2020nvy}.
The concerned energy region is also sensitive to the promising DUNE experiment, which is expected to observe 
significant $\nu_\tau$ events. Since the liquid argon time projection chamber~\cite{DUNE:2020ypp}
has been recently considered to be the most promising detector to observe neutrino events, therefore, 
we have performed the present calculations for the $^{40}Ar$ nucleus treating it to be an isoscalar nuclear target. The scattering cross section and the
polarization observables $P_L$ and $P_T$ are expressed in terms of the nuclear structure functions $F_{iA}(x,Q^2);~(i=1-5)$. These nuclear structure functions are convolution of the nucleon spectral function
and the nucleon structure functions $F_{iN}(x,Q^2);~(i=1-5)$. Various perturbative and non-perturbative effects which modify the nucleon structure functions are taken into account
in the different regions of $x$ and $Q^2$. The special features of this work are:

\begin{itemize}
 \item The nucleon structure functions are calculated at NLO and the $Q^2$ evolution of parton densities is performed by using the formalism developed in Ref.~\cite{Ansari:2020xne, Kretzer:2002fr}
 in calculating the structure functions beyond the leading order.
 \item TMC in evaluating the nucleon structure functions has been taken into account following the works of Kretzer et al.~\cite{Kretzer:2003iu}.
 \item These nucleon structure functions have been used to calculate the nuclear structure functions.
 \item The effects of the nuclear medium have been studied in a microscopic field theoretical model that uses relativistic
nucleon spectral function incorporating the effects of Fermi
motion, binding energy, and nucleon correlations~\cite{FernandezdeCordoba:1991wf}. The nucleon spectral function describes the energy and momentum distribution of the nucleons in
the nucleus and is obtained by using the Lehmann's representation for the relativistic nucleon propagator and nuclear many-body
theory is used to calculate it for an interacting Fermi sea in the nuclear medium~\cite{Marco:1995vb, FernandezdeCordoba:1991wf}.
The numerical calculations are performed in the local density approximation, where density is a function of point of the interaction
 $\vec r$ for a given volume element. 
\end{itemize}

 In section~\ref{formalism}, we present the formalism for the unpolarized and polarized $\tau-$lepton in the charged current $\nu_\tau/\bar\nu_\tau$ induced DIS
 processes off the free nucleon ($N$) and the nuclear ($A$) targets. In Sec.~\ref{res}, the results are presented followed by discussions and conclusions.
\section{Formalism}\label{formalism}
\subsection{$\nu_\tau-N$ DIS: unpolarized $\tau-$lepton}
\begin{figure}[h]
\begin{center}
\includegraphics[height=5 cm, width=7 cm]{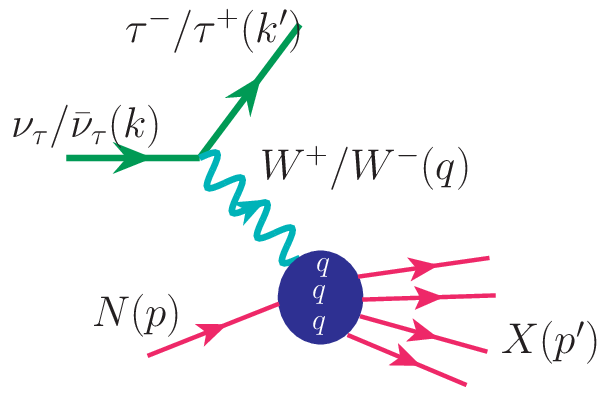}
\end{center}
\caption{Feynman diagram for the $\nu_\tau/\bar\nu_\tau$ induced DIS process off free nucleon target ($N$).}
\label{fig_fyn0}
\end{figure}

\begin{figure}[h]
 \includegraphics[height=7.0 cm, width=0.9\textwidth]{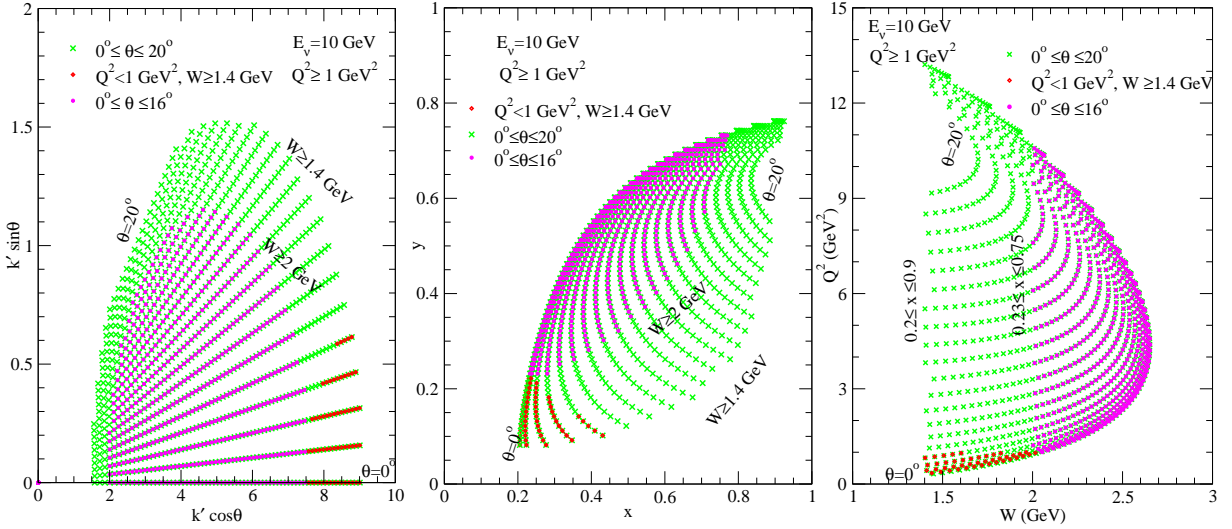}
\caption{Kinematically allowed region for $\nu_\tau-N$ induced DIS process at $E_{\nu_\tau}=10$ GeV with the different constrains on center of mass energy $W$, i.e. $W\ge 1.4$ GeV and $W\ge 2$ GeV. 
Diamond shaped points corresponds to the region where $Q^2<1$ GeV$^2$. }
\label{fig0}
\end{figure}

The charged current $\nu_\tau/\bar\nu_\tau-N$ deep 
inelastic scattering process depicted in Fig.~\ref{fig_fyn0} is 
\begin{equation}
 \nu_\tau/\bar\nu_\tau(k)+N(p) \rightarrow \tau^-/\tau^+(k')+X(p'),
\end{equation}
where inside the parenthesis four momenta of the corresponding particles are written. 
In the laboratory frame, where the target nucleon is at rest, the four momenta are written as
\begin{eqnarray}
 k^\mu=(k^0,\vec k)&=&(E_{\nu_\tau},0,0,E_{\nu_\tau}),\nonumber\\
 k'^\mu=(k'^0,\vec k')&=&(E_\tau, |\vec k'| \sin\theta,0,|\vec k'| \cos\theta),\nonumber\\
 p^\mu=(p^0,\vec p)&=&(M_N,0,0,0),\nonumber
\end{eqnarray}
where $|\vec k'|=\sqrt{E_\tau^2-m_\tau^2}$ and $m_\tau$ is the outgoing $\tau-$lepton.
The expression of differential scattering cross section for the unpolarized $\tau-$lepton production is given by~\cite{Haider:2016zrk}:
 \begin{eqnarray}\label{xsec0}
  \frac{d^2\sigma_{N}}{dE_\tau d\cos\theta}&=&\frac{G_F^2 |\vec k'|}{2\pi E_{\nu_\tau}\left(1+\frac{Q^2}{M_W^2}\right)^2}  L_{\mu\nu}\;W_{N}^{\mu \nu},
\end{eqnarray}
with four momentum transfer square $Q^2=-q^2=(k-k')^2$, $G_F$ is the Fermi coupling constant and $M_W$ is the mass of intermediate vector W-boson.
$W_{N}^{\mu \nu}$ is the nucleon hadronic tensor given by~\cite{Ansari:2020xne}:
 \begin{eqnarray}
 \label{nucleon_had_ten_weak}
W_{N}^{\mu \nu} &=&
 -g^{\mu \nu}\;W_{1N} (\nu, Q^2)
+ \frac{p^{\mu}\;p^{\nu}}{M_N^2}W_{2N} (\nu, Q^2)
-\frac{i}{M_N^2} \epsilon^{\mu \nu \rho \sigma} p_{ \rho} q_{\sigma}\;
W_{3N} (\nu, Q^2) + \frac{q^{\mu} q^{\nu}}{M_N^2} \;W_{4N} (\nu, Q^2)\nonumber\\
&&+\frac{(p^{\mu} q^{\nu} + q^{\mu} p^{\nu})}{M_N^2} \;W_{5N} (\nu, Q^2)
+ \frac{i (p^{\mu} q^{\nu} - q^{\mu} p^{\nu})}{M_N^2} 
W_{6N} (\nu, Q^2)\,,
\end{eqnarray}
and the unpolarized leptonic tensor $L_{\mu\nu}$ is given by~\cite{Ansari:2020xne}:
\begin{equation}\label{lep_ten}
 L_{\mu\nu}=8(k_\mu k'_\nu+k_\nu k'_\mu-k\cdot k' g_{\mu\nu}\pm i\epsilon_{\mu\nu\rho\sigma} k^\rho k'^\sigma),
\end{equation}
``+'' sign is for antineutrino and ``-'' sign is for neutrino.
Using Eqs.~\ref{nucleon_had_ten_weak} and \ref{lep_ten} in Eq.~\ref{xsec0}, we obtain the following expression of unpolarized double differential scattering cross section
\begin{eqnarray}\label{xsec1}
 \frac{d^2\sigma_N}{dE_\tau\;d\cos\theta}&=& \frac{G_F^2 |\vec k'|}{2\pi (1+\frac{Q^2}{M_W^2})^2}\;\left[ 2 W_{1N}(\nu, Q^2) (E_\tau-|\vec k'| \cos\theta)+ 
 W_{2N}(\nu, Q^2)\;(E_\tau+|\vec k'|\cos\theta)\right.\nonumber\\
 &\pm& \left.W_{3N} (\nu, Q^2)\;\frac{2}{M_N}(|\vec k'|^2+E_{\nu_\tau} E_\tau-(E_{\nu_\tau}+E_\tau)|\vec k'|\;\cos\theta)+W_{4N}(\nu, Q^2)\;\frac{m_\tau^2}{M_N^2}(E_\tau-|\vec k'| \cos\theta)\right.\nonumber\\
 &-&\left.W_{5N}(\nu, Q^2)\;\frac{2 m_\tau^2}{M_N}\right],
\end{eqnarray}
with upper sign for neutrino and lower sign for antineutrino.
The contribution of the term with $W_{6N} (\nu, Q^2)$ vanishes when contracted with the leptonic tensor $ L_{\mu\nu}$. When $Q^2$ and $\nu$ become 
large the structure functions $W_{iN}  (\nu,Q^2);~(i=1-5)$ are generally expressed in terms of the dimensionless nucleon structure functions $F_{iN}  (x),\;\;i=1 - 5$ as: 
 \begin{eqnarray}
 \label{ch2:relation}
 F_{1N}(x) &=& W_{1N}(\nu,Q^2) ;\hspace{3 mm} F_{2N}(x) = \frac{Q^2}{2xM_N^2}W_{2N}(\nu,Q^2); \hspace{3mm}  F_{3N}(x) = \frac{Q^2}{xM_N^2}W_{3N}(\nu,Q^2)\\
 F_{4N}(x) &=& \frac{Q^2}{2M_N^2}W_{4N}(\nu,Q^2) ;\hspace{3 mm} F_{5N}(x) = \frac{Q^2}{2xM_N^2}W_{5N}(\nu,Q^2). 
   \label{ch3:relation}
\end{eqnarray}
We observe that the scattering cross section (Eq.~\ref{xsec1}) obtained using the aforementioned relations does not satisfy the positivity condition in neighborhood of the threshold region. Therefore, 
we made the following modification in $F_{1N}(x)$ structure function in order to ensure the positivity constraints as a result of which the allowed kinematic region for the 
$\tau-$lepton polarization get reduced~\cite{Hagiwara:2003di, Bourrely:2004iy}:
\begin{equation}
 F_{1N}(x)=\frac{W_{1N}(\nu,Q^2)}{\left(1+\frac{x M_N}{\nu}\right)}.
\end{equation}
In the above expressions, $x\big(=\frac{Q^2}{2 p \cdot q}\big)$  is the Bjorken scaling variable which lie in the range of $ \frac{m_\tau^2}{2M_N (E_{\nu_\tau} - m_\tau)} \le x \le 1$ and $\nu=y E_{\nu_\tau}=E_{\nu_\tau}-E_\tau$ is the 
 energy transfer. In Fig.~\ref{fig0}, we have shown the allowed kinematic region in the ``$|\vec k'| \cos\theta$''-``$|\vec k'|\sin\theta$'' plane (left panel), 
 ``Bjorken $x$''-`` inelasticity $y$'' plane (middle panel), and ``$W$''-``$Q^2$'' plane (right-panel), at $E_{\nu_\tau}=10$ GeV. 
 It may be noticed that 
 the kinematic region with $W\ge 2$ GeV cut is much suppressed as compared to the case when a cut of $W\ge 1.4$ GeV is applied. The existing ambiguity in the choice of center
 of mass energy cut in the literature needs more attention from the particle physics community and has been recently discussed in Ref.~\cite{SajjadAthar:2020nvy}. We have considered the region of $Q^2\ge 1$ GeV$^2$ and $W\ge 2$ GeV to 
 be the region of true DIS events and apply
 these constrains in the numerical calculations. It must be noticed that for the true DIS events we could not dig out the information from the regions of low as well as high values of 
 $x$ and $y$, only the intermediate region of scaling variables ($ 0.2\le (x,y) <0.8$) is accessible 
 at the chosen value of $\nu_\tau$ energies in the present work. 

Using the relations given in Eqs.~\ref{ch2:relation} and \ref{ch3:relation} in Eq.~\ref{xsec1}, we obtain the differential scattering cross section 
in terms of the dimensionless nucleon structure functions
\begin{eqnarray}\label{xsec2}
 \frac{d^2\sigma_N}{dE_\tau\;d\cos\theta}&=& \frac{G_F^2 |\vec k'|}{2\pi M_N (1+\frac{Q^2}{M_W^2})^2}\;\left[ 2 F_{1N}(x, Q^2) (E_\tau-|\vec k'| \cos\theta)+
 F_{2N}(x, Q^2)\;\frac{M_N}{\nu}(E_\tau+|\vec k'|\cos\theta)\right.\nonumber\\
 &\pm& \left.F_{3N} (x, Q^2)\;\frac{1}{\nu}(|\vec k'|^2+E_{\nu_\tau} E_\tau-(E_{\nu_\tau}+E_\tau)|\vec k'|\;\cos\theta)+F_{4N}(x, Q^2)\;\frac{m_\tau^2}{\nu M_N x}(E_\tau-|\vec k'| \cos\theta)\right.\nonumber\\
 &-&\left.F_{5N}(x, Q^2)\;\frac{2 m_\tau^2}{\nu}\right],
\end{eqnarray}
The dimensionless nucleon structure functions are written in terms of the parton distribution functions $q_i(x)$ and $\bar q_i(x)$ at the leading order as
\begin{eqnarray}\label{parton_wk}
F_{2N}(x)  &=& \sum_{i} x [q_i(x) +\bar q_i(x)] \;;\;
x F_{3N}(x) =  \sum_i x [q_i(x) -\bar q_i(x)]\;;\;\;
F_{4N}(x)=0.
\end{eqnarray} 
 For an isoscalar nucleon target $F_{2N}(x)$ and $F_{3N}(x)$ in the four flavor scheme are given by:
\begin{eqnarray}\label{eq:pdf1}
F_{2N}^{\nu_\tau}(x)& = &  x [u(x)+d(x) +\bar u(x)+\bar d(x)+2 s(x) +2\bar{c}(x)],\;\;\;F_{2N}^{\bar\nu_\tau}(x)=  x [u(x)+d(x) +\bar u(x)+\bar d(x)+2 \bar s(x) +2{c}(x)]\nonumber\\
x F_{3N}^{\nu_\tau}(x)& = &  x [u(x)+d(x) - \bar{u}(x) -\bar{d}(x)+2 s(x)-2\bar c(x)],\;\;\;x F_{3N}^{\bar\nu_\tau}(x)=  x [u(x)+d(x) - \bar{u}(x) -\bar{d}(x)-2 \bar s(x)+2 c(x)]\nonumber
\end{eqnarray}
We have treated all the four flavors of quarks to be massless as we have found that the effect of massive charm quark is very small in the present kinematic region.
These details have already been discussed in Refs.~\cite{Ansari:2020xne, Zaidi:2021iam}.
To define the structure functions $F_{1N}(x)$ and $F_{5N}(x)$ at the leading order we use Callan-Gross~\cite{Callan:1969uq} and Albright-Jarlskog~\cite{Albright:1974ts} relations, respectively as:
\begin{eqnarray}
 F_{1N}(x)&=&\frac{F_{2N}(x)}{2 x}\;;\;\;
 F_{5N}(x)=\frac{F_{2N}(x)}{2 x} \nonumber
\end{eqnarray}
As the kinematic region of the low and moderate
 $Q^2$, the region of our particular interest, is sensitive to the perturbative and non-perturbative QCD corrections such as the evolution of parton densities beyond the leading order
 and the TMC effect, respectively, these effects have been taken into account while evaluating the nucleon structure functions~\cite{Ansari:2020xne}. We have
 used the Martin-Motylinski-Harland-Lang-Thorne (MMHT) parton distribution functions (PDFs) parameterization~\cite{Harland-Lang:2014zoa} up to NLO in the
4-flavor (u; d; s, and c) minimal subtraction (MSbar) scheme.
Furthermore, target mass correction effect has been incorporated following Refs.~\cite{Ansari:2020xne, Kretzer:2003iu}.
 
 For the polarized tau-lepton, the hadronic current remains the same while the leptonic current will gets modified (see Fig.~\ref{fig_fyn0}). In the next section, we present the formalism for the 
 production of polarized $\tau-$lepton in the 
 charged current induced $\nu_\tau/\bar\nu_\tau-N$ DIS processes.
\subsection{$\nu_\tau-N$ DIS: polarized $\tau-$lepton}
\begin{figure}[h]
\begin{center}
\includegraphics[height=6 cm, width=6 cm]{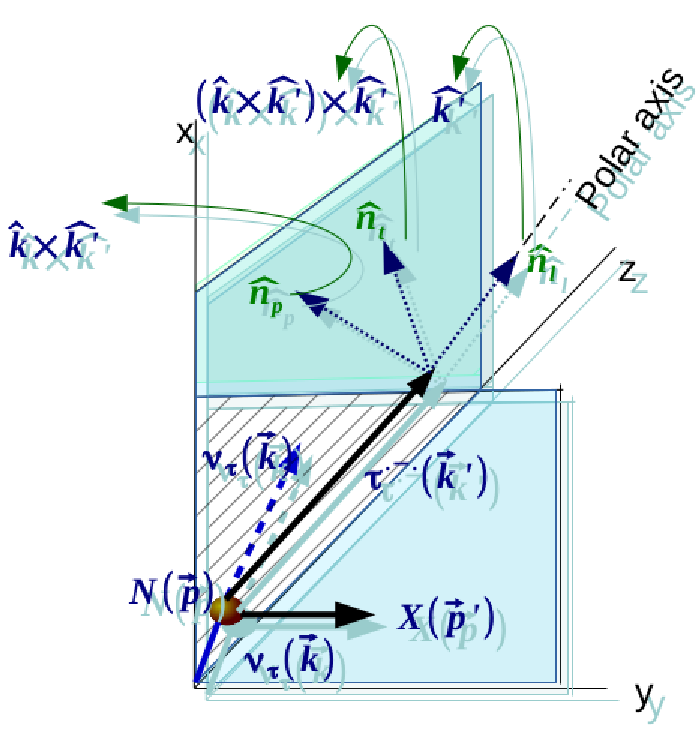}
\end{center}
\caption{ Momentum and polarization directions of the produced $\tau-$lepton. The orthogonal unit vectors $\hat n_l$, $\hat n_t$, and $\hat n_{p}$, respectively correspond to the 
longitudinal, transverse and perpendicular
directions with respect to the $\tau-$lepton momentum.}
\label{rot_frame}
\end{figure}

In the case of polarized $\tau^\mp-$lepton (as depicted in Fig.~\ref{rot_frame}) the leptonic tensor modifies to~\cite{Sobczyk:2019urm}
\begin{equation}\label{lep_pol}
 L_{\mu\nu}(s;h)=\frac{1}{2}[L_{\mu\nu} \mp h\;m_\tau s^\alpha\;(k_\mu g_{\nu\alpha}+k_\nu g_{\mu\alpha}-k_\alpha g_{\mu\nu}\pm i \;\epsilon_{\mu\nu\alpha\beta} k^\beta)],
\end{equation}
where $h=\pm 1$ is the helicity, $s^\alpha$ is the spin four-vector, and upper sign is for neutrino and lower sign is for antineutrino. We choose the axis of quantization specified by the unit vector $\hat n$ in the rest frame of the $\tau-$lepton 
($k'^\mu=(k^0,\vec 0)$) and define $s^\alpha$ in this frame as follows
\begin{equation}
 s^\alpha=(0,\hat n)
\end{equation}
However, in any other frame it is given by~\cite{Sobczyk:2019urm}:
\begin{equation}
 s^\alpha=\left(\frac{\vec k' \cdot \hat n}{m_\tau},\hat n+\frac{\vec k'}{m_\tau}\;\frac{(\vec k' \cdot \hat n)}{(E_\tau+m_\tau)} \right),
\end{equation} 
following the scalar product invariance and ensuring that $s^2=-1$ and $s \cdot k'=0$. Using these expressions we have obtained the polarized double differential scattering cross section as
\begin{eqnarray}
 \frac{d^2\sigma_{N}^{pol}}{dE_\tau\;d\cos\theta}&=&\frac{G_F^2 |\vec k'|}{2\pi E_{\nu_\tau}(1+\frac{Q^2}{M_W^2})^2}  L_{\mu\nu}(s;h)\;W_{N}^{\mu \nu}
 \end{eqnarray}
 After contracting the polarized leptonic tensor given in Eq.~\ref{lep_pol} with the hadronic tensor (Eq.~\ref{nucleon_had_ten_weak}), we get~\cite{Sobczyk:2019urm, Hernandez:2022nmp}
 \begin{eqnarray}
 \frac{d^2\sigma_{N}^{pol}}{dE_\tau\;d\cos\theta}&=&(1\pm h s^\alpha {\cal P}^\alpha)\frac{d^2\sigma_N}{dE_\tau\;d\cos\theta},
\end{eqnarray}
where the upper sign stands for the neutrino, the lower sign is for the antineutrino induced processes, and ${\cal P}^\alpha$ is the polarization vector of the final state charged lepton that is obtained
by taking the nucleon response~\cite{Sobczyk:2019urm, Hernandez:2022nmp}:
\begin{equation}
 {\cal P}^\alpha=\mp\frac{m_\tau\;(k_\mu g_{\nu\alpha}+k_\nu g_{\mu\alpha}-k_\alpha g_{\mu\nu}\pm i \;\epsilon_{\mu\nu\alpha\beta} k^\beta)\;W_{N}^{\mu \nu}}{L_{\mu\nu}\;W_{N}^{\mu \nu}}.
\end{equation}
The polarization vector is decomposed into longitudinal (in the direction to the outgoing $\tau-$lepton momentum $\vec k'$), 
perpendicular (in the orthogonal direction to $\nu_\tau-\tau$ plane) and transverse (transverse to $\vec k'$ and in the $\nu_\tau-\tau$ plane) components as:
\begin{equation}
  {\cal P}^\alpha=-(\underbrace{P_L\;n_l^\alpha}_{longitudinal}+\underbrace{P_T\;n_t^\alpha}_{transverse}+\underbrace{P_{P}\;n_{p}^\alpha}_{perpendicular}),
\end{equation}
where $n^\alpha_i; ~(i=l,t,p)$ are the orthogonal basis of the four-vector Minkowski space, and are defined as
\begin{equation}
n_l^\alpha=\left(\frac{|\vec k'|}{m_\tau}, \frac{E_\tau \vec k'}{m_\tau |\vec k'|} \right),\hspace{3 mm} n_t^\alpha=\left(0,\frac{(\vec k \times \vec k')\times \vec k'}{|(\vec k \times \vec k')\times \vec k'|} \right),\hspace{3 mm} n_{p}^\alpha=\left(0,\frac{\vec k \times \vec k'}{|\vec k \times \vec k'|} \right)
\end{equation}
Thus the different polarization components are obtained as
\begin{equation}
 P_L=-{\cal P}^\alpha \cdot n_l^\alpha, \hspace{3 mm} P_T=-{\cal P}^\alpha \cdot n_t^\alpha, \hspace{3 mm} P_{P}=-{\cal P}^\alpha \cdot n_{p}^\alpha.
\end{equation}
We find that ${\cal P}^\alpha \cdot n_{p}^\alpha=0$, which implies that the perpendicular component ($P_{P}$) does not contribute as the polarization three-vector
lies in the direction perpendicular to the $\tau-$lepton scattering plane. However, on simplification of the above expression in the laboratory frame, we obtain ~\cite{Hagiwara:2003di, Hernandez:2022nmp}
\begin{eqnarray}\label{polcomp}
  P_L&=&\mp\frac{E_{\nu_\tau}}{L_{\mu\nu}\;W_{N}^{\mu \nu}}\left[\Big(2 F_{1N}(x,Q^2)-\frac{m_\tau^2}{M_N \nu x} F_{4N}(x,Q^2) \Big)(|\vec k'|-E_\tau\;\cos\theta)+F_{2N}(x,Q^2) \frac{M_N}{\nu} (|\vec k'|+E_\tau \;\cos\theta)\right.\nonumber\\
  &&\left.\pm\frac{F_{3N}(x,Q^2)}{\nu}(|\vec k'|(E_{\nu_\tau}+E_\tau)-(|\vec k'|^2+E_{\nu_\tau} E_\tau)\;\cos\theta) -\frac{2}{\nu} m_\tau^2 \;\cos\theta\;F_{5N}(x,Q^2)\right],\\
  P_T&=&\mp\frac{m_\tau\;sin\theta\;E_{\nu_\tau}}{L_{\mu\nu}\;W_{N}^{\mu \nu}}\left[2 F_{1N}(x,Q^2)- F_{2N}(x,Q^2) \frac{M_N}{\nu}\pm \frac{E_{\nu_\tau}}{\nu}\;F_{3N}(x,Q^2)-\frac{m_\tau^2}{M_N \nu x} F_{4N}(x,Q^2)\right.\nonumber\\
  &&\left.+\frac{2 E_\tau}{\nu} \;F_{5N}(x,Q^2)\right],
\end{eqnarray}
where the upper sign corresponds to the case of neutrino and the lower sign corresponds to the antineutrino case. It may be noticed that the polarization components depend upon the lepton kinematic variables, i.e. the 
lepton energy $E_\tau$ and scattering angle $\theta$ as well as upon the dimensionless nucleon structure functions $F_{iN}(x,Q^2);~(i=1-5)$, and thus exhibit the particular sensitivities to the 
different elements used in the scattering cross section models.
The degree of polarization $P$, a Lorentz invariant quantity, is given by~\cite{Hagiwara:2003di}
\begin{equation}
 P^2=(P_L^2+P_T^2)  \Rightarrow P=\sqrt{P_L^2+P_T^2} \le 1
\end{equation}
For a fully polarized $\tau-$lepton, $P=1$, while in the case of 
the partially polarized $\tau-$lepton the degree of polarization lies in the range of 
$0< P < 1$, and for the unpolarized $\tau-$lepton the degree of polarization is zero. To specify the direction of polarization we define the spin polarization vector in the rest
frame of the $\tau-$lepton as~\cite{Hagiwara:2003di}
\begin{equation}\label{spin_comp}
 \vec s=(s_x,s_y,s_z)=P(\sin\theta_P\;\cos\phi_P,\sin\theta_P\;\sin\phi_P,\cos\theta_P),
\end{equation}
where the $z-$axis is taken along the direction of tauon momentum, $\theta_P$ is the polar angle and $\phi_P$ is the azimuthal angle of the spin polarization vector. $\phi_P$ may have values either 0 or $\pi$ 
corresponding to $\vec k'$, constraining that the polarization vector always remain in the scattering plane. Hence the direction of polarization is determined by the polar component of normalized polarization vector, i.e.~\cite{Hagiwara:2003di}
\begin{equation}\label{dircos}
 \cos\theta_P=\frac{s_z}{P}=\frac{P_L}{P}
\end{equation}

\subsection{$\nu_\tau-A$ DIS: Nuclear medium effects }
\begin{figure}
\begin{center}
\includegraphics[height=5 cm, width=7 cm]{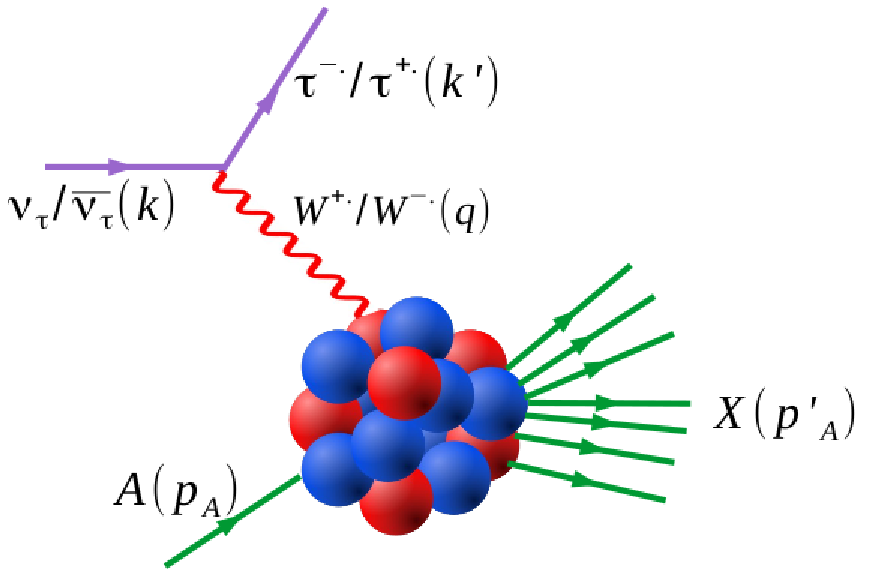}
\end{center}
\caption{Feynman diagrams for the $ \nu_\tau / \bar\nu_\tau$ induced DIS process off nuclear target ($A$).}
\label{dis_nuc}
\end{figure}

 The differential scattering cross section for the charged current inclusive $ \nu_\tau / \bar\nu_\tau$-nucleus deep inelastic scattering process (depicted in Fig.~\ref{dis_nuc}):
\begin{equation}\label{DISrxna}
 \nu_\tau / \bar\nu_\tau(k) + A(p_A) \rightarrow \tau^-/\tau^+(k') + X(p'_A)
\end{equation}
 is given by
  \begin{eqnarray}\label{xsec0a}
  \frac{d^2\sigma_{A}}{dE_\tau d\cos\theta}&=&\frac{G_F^2 |\vec k'|}{2\pi E_{\nu_\tau}(1+\frac{Q^2}{M_W^2})^2}  L_{\mu\nu}\;W_{A}^{\mu \nu},
\end{eqnarray}
where $W_{A}^{\mu \nu}$ is the nuclear hadronic tensor and is expressed in terms of the dimensionless nuclear structure functions $F_{iA}(x);~(i=1-5)$ as
 \begin{eqnarray}
 \label{nuc_had_weak}
W_{A}^{\mu \nu} &=&
 -g^{\mu \nu}\;W_{1A} (\nu, Q^2)
+ \frac{p^{\mu}_A\;p^{\nu}_A}{M_N^2}W_{2A} (\nu, Q^2)
-\frac{i}{M_N^2} \epsilon^{\mu \nu \rho \sigma} p_{A \rho} q_{\sigma}\;
W_{3A} (\nu, Q^2) + \frac{q^{\mu} q^{\nu}}{M_N^2} \;W_{4A} (\nu, Q^2)\nonumber\\
&&+\frac{(p^{\mu}_A q^{\nu} + q^{\mu} p^{\nu}_A)}{M_N^2} \;W_{5A} (\nu, Q^2)
+ \frac{i (p^{\mu}_A q^{\nu} - q^{\mu} p^{\nu}_A)}{M_N^2} 
W_{6A} (\nu, Q^2)\,,
\end{eqnarray}
Following the same analogy as that for a free nucleon target, we have obtained the following expression of the double differential scattering cross section
for the nucleons bound inside a nuclear target:
\begin{eqnarray}\label{xsecA}
 \frac{d^2\sigma_A}{dE_\tau\;d\cos\theta}&=& \frac{G_F^2 |\vec k'|}{2\pi M_N (1+\frac{Q^2}{M_W^2})^2}\;\left[ 2 F_{1A}(x, Q^2) (E_\tau-|\vec k'| \cos\theta)+ F_{2A}(x, Q^2)\;\frac{M_N}{\nu}(E_\tau+|\vec k'|\cos\theta)\right.\nonumber\\
 && \left.\pm \;F_{3A} (x, Q^2)\;\frac{1}{\nu}(|\vec k'|^2+E_{\nu_\tau} E_\tau-(E_{\nu_\tau}+E_\tau)|\vec k'|\;\cos\theta)+F_{4A}(x, Q^2)\;\frac{m_\tau^2}{\nu M_N x}(E_\tau-|\vec k'| \cos\theta)\right.\nonumber\\
 &&\left.-\;F_{5A}(x, Q^2)\;\frac{2 m_\tau^2}{\nu}\right]
\end{eqnarray}
In the present work, we consider the scattering process in the laboratory frame, where the target nucleus is at rest i.e. $p_A=(p_A^0,~{\vec p_A}= 0)$ and the momentum of the nucleon (${\vec p_N}$)
in the nucleus is non-zero, and the motion of such nucleons corresponds to the 
Fermi motion. In the local density approximation, the lepton scatters from a bound nucleon having density $\rho_N(r)$ and the corresponding Fermi
momentum $p_{F_N}(r)=(3 \pi^2 \rho_N(r))^{1/3}$. The local density $\rho_N(r)$ of protons or neutrons bound inside the nucleus is given by $\rho_p(r)=\frac{Z}{A}\rho(r)$ and 
$\rho_n(r)=\frac{A-Z}{A}\rho(r)$, where $\rho(r)$ is the nuclear density which is calculated by using the parameterization of 2pF density, i.e.,
$\rho(r)=\frac{\rho_0}{1+e^{(r-c_1) / c_2}}$. $\rho_0$ is the central density, and $c_1$ and $c_2$ are the density parameters which are 
determined through the electron scattering experiments~\cite{DeJager:1974liz, DeVries:1987atn}. The details can be found in Ref.~\cite{AtharSajjad:2022ipr}. The differential cross
section ($d\sigma_A$) for $\nu_\tau/\bar\nu_\tau-A$ scattering is obtained by folding $\nu_\tau/\bar\nu_\tau-N$
scattering cross section ($d\sigma_N$)
with $\rho_N(r)$ and integrating over the volume element:
\begin{equation}
 d\sigma_A=\int d^3r\;\rho_N(r)\;d\sigma_N, 
\end{equation}
In the case of a symmetric nuclear matter, each nucleon occupies a volume of $(2\pi \hbar)^3$, and due to the two possible spin 
orientations of the nucleon, each unit cell in the configuration space
is occupied by the two nucleons. Hence, the number of nucleons for a given volume ($\hbar=1$ in natural units):
\begin{eqnarray}
 N &=& 4 V \;\int_0^{p_{F_N}}\;\frac{d^3p_N}{(2 \pi)^3}\; n({\vec p_N},\vec r),\\
 \Rightarrow \rho &=& {N \over V} = 4 \int_0^{p_{F_N}}\;\frac{d^3p_N}{(2 \pi)^3}\; n({\vec p_N}, \vec r),
\end{eqnarray}
where $n({\vec p_N},\vec r)=\theta(p_{F_{N}}(\vec r)-|\vec p_N|)$ is the occupation number of initial nucleon carrying momentum $\vec p_N$, and it has dependence on the 
radial coordinates due to the Fermi momentum $p_{F_N}(r)$. The occupation number corresponding to the nucleon within the Fermi sea, fulfills the following condition:
\begin{equation}
n({\vec p_N}) = 
\left\{
 \begin{array}{c}
1 \;\;\;\mbox{for ${\vec p_N} \le {\vec p_{F_N}}$}  \\
0 \;\;\;\mbox{for ${\vec p_N} > {\vec p_{F_N}}$}
 \end{array}\right..
\end{equation}
We have chosen the four momentum transfer along the $z$-axis, therefore, $q^\mu=(q^0,0,0,q^z)$ and the Bjorken variable $x_N$ corresponding to the nucleon bound inside a nucleus is 
then expressed as:
 \begin{equation}
 x_N = \frac{Q^2}{2 p_N \cdot q} = \frac{Q^2}{2 (p_N^0 q^0 - p_N^z q^z)}.
\end{equation}
The Bjorken variable for the nuclear target $x_A$ is given by:
\begin{eqnarray}
x_A&=&\frac{Q^2}{2 p_A \cdot q}=\frac{Q^2}{2 p_{A}^0  q^0 } = \frac{Q^2}{2 A~M_N q^0}=\frac{x}{A}.
\end{eqnarray}
The nucleons inside a nuclear target interact among themselves via the strong interaction, hence various nuclear medium effects come into play depending upon the 
value of the Bjorken variable $x$. The nuclear effects such as Fermi motion, binding and nucleon correlations have been taken into account for an inclusive process by using the 
hole spectral function ($S_h$) calculated in a microscopic field theoretical model~\cite{FernandezdeCordoba:1991wf, Marco:1995vb}. The hole spectral function is properly normalized and has been
checked by obtaining the correct baryon number ($A$) and nuclear binding energy for a given nuclear target~\cite{Zaidi:2019asc, Haider:2015vea}:
\begin{equation}\label{shnorm}
4\int\;d^3r\;\int\frac{d^3 p_N}{(2\pi)^3}\;\int_{-\infty}^\mu \;S_h(\omega, \vec{p_N},\rho(r))\;d\omega=A,
\end{equation}
where $\mu$ is the chemical potential and $\omega$ is the removal energy~\cite{Marco:1995vb}.
We have obtained the differential scattering cross section
in terms of the hole spectral function as~\cite{Zaidi:2021iam, AtharSajjad:2022ipr}:
\begin{eqnarray}\label{conv_xseca}
\frac{d^2\sigma_A}{dE_\tau\;d\cos\theta}&=&\frac{G_F^2 |\vec k'|}{2\pi E_{\nu_\tau}\left(1+{Q^2\over M_W^2} \right)^2}\;L_{\mu\nu} 4 \int \, d^3 r \, \int \frac{d^3 p_N}{(2 \pi)^3} \, 
\frac{M_N}{E ({\vec p_N})} \, \int^{\mu}_{- \infty} d p_N^0 S_h (p_N^0, {\vec p_N}, \rho(r))
W^{\mu \nu}_{N} (p_N, q), \,
\end{eqnarray}
Comparing Eqs.~\ref{xsec0a} and \ref{conv_xseca}, we obtain the nuclear hadronic tensor $W^{\mu \nu}_{A}$ 
 in terms of the nucleonic tensor $W^{\mu \nu}_{N}$ convoluted over the hole spectral function $S_h$~\cite{Zaidi:2021iam}:
 \begin{equation}\label{conv_WAa}
W^{\mu \nu}_{A} = 4 \int \, d^3 r \, \int \frac{d^3 p_N}{(2 \pi)^3} \, 
\frac{M_N}{E ({\vec p_N})} \, \int^{\mu}_{- \infty} d p_N^0 S_h (p_N^0, {\vec p_N}, \rho(r))
W^{\mu \nu}_{N} (p_N, q), \,
\end{equation}
 where the factor of 4 is because of the spin-isospin degrees of freedom of the nucleon for an isoscalar nuclear target. 
We take the appropriate components of the nucleon ($W^{\mu\nu}_N$ in Eq.\ref{nucleon_had_ten_weak}) 
and the nuclear ($W^{\mu\nu}_A$ in Eq.\ref{nuc_had_weak}) hadronic tensors along 
the $x$, $y$ and $z$ axes and obtained the expressions for the nuclear structure functions as~\cite{Zaidi:2021iam}: 
  \begin{eqnarray}\label{spect_funct}
F_{iA} (x_A, Q^2) &=& 4\int \, d^3 r \, \int \frac{d^3 p_N}{(2 \pi)^3} \, 
\frac{M_N}{E_N ({\vec p_N})} \, \int^{\mu}_{- \infty} d p^0~ S_h(p_N^0, {\vec p_N}, \rho(r))~
\times f_{iN}(x,Q^2);~~~~~i=1-5,
\end{eqnarray}
where 
\begin{eqnarray}
 f_{1N}(x,Q^2)&=&AM_N\left[\frac{F_{1N} (x_N, Q^2)}{M_N} + \left(\frac{p_N^x}{M_N}\right)^2 \frac{F_{2N} (x_N, Q^2)}{\nu_N}\right],\\
f_{2N}(x,Q^2)&=&\left( \frac{F_{2N}(x_N,Q^2)}{M_N^2 \nu_N}\right)\left[ \frac{Q^4}{q^0 {(q^z)}^2}\left(p_N^z+\frac{q^0(p_N^0-\gamma p_N^z) }{Q^2}{ q^z} \right)^2+\frac{q^0 Q^2 (p_N^x)^2}{{(q^z)}^2}\right]\\
f_{3N}(x,Q^2)&=&A\frac{q^0}{q^z} \;\times\left({p_N^0 q^z - p_N^z q^0  \over p_N \cdot q} \right)F_{3N} (x_N,Q^2),\\
f_{4N}(x,Q^2)&=&A\; \left[F_{4N}(x_N,Q^2) +\frac{p_N^z Q^2}{{ q^z}} \frac{F_{5N}(x,Q^2)}{M_N \nu_N}\right],\\
f_{5N}(x,Q^2)&=&A\;\;\frac{F_{5N}(x_N,Q^2)}{M_N \nu_N}\times\left[q^0(p_N^0-\gamma p_N^z)+Q^2 \frac{p_N^z}{{ q^z}} \right]
\end{eqnarray}
and $\gamma=\frac{q^z}{q^0}$.

The nucleons which are bound inside the nucleus interact with each other via meson exchange such as 
$\pi,~\rho,$ etc. The interaction of the intermediate vector boson (IVB) 
 with the mesons play an important role in the evaluation of nuclear structure functions~\cite{Haider:2012nf,Haider:2015vea, Haider:2016zrk, Zaidi:2019asc, Zaidi:2019mfd, Zaidi:2021iam}. 
 In the low region of $x$ the contribution from shadowing ($0<x< 0.1$) and antishadowing ($0.1<x<0.2$) effects is important. However, in the present kinematic region of $x$ and $Q^2$, we have
 found that the mesonic contribution and the (anti)shadowing effect are almost negligible, therefore, we have not included these effects while presenting the results. It must be pointed out that in our earlier studies we have 
observed the non-isoscalarity corrections in argon to be very small in the kinematic region of present interest~\cite{Ansari:2021cao, Zaidi:2021iam, Zaidi:2022qdp, AtharSajjad:2022ipr}.
All the numerical results are obtained in the energy range of $6\le E_{\nu_\tau} \le 20$ GeV by taking into account the target mass correction effect at the next-to-leading order
in the kinematic region of $Q^2\ge 1$ GeV$^2$. To understand the effect of the center of mass energy cut we have obtained the results in various kinematic regions using different
cuts on $W$, i.e., $W\ge 1.4$ GeV, 
$W \ge 1.6$ GeV and $W \ge 2$ GeV.

\section{Discussion of Results and Conclusions}\label{res}
In this section, we present the results for the $\tau-$lepton production cross section by using Eqs.~\ref{xsec2} and \ref{xsecA} as well as its polarization observables such as the degree of
polarization ($P$), longitudinally ($P_L$),
and transversely ($P_T$) polarized components and the direction of polarization vector as defined through Eqs.~\ref{polcomp}-\ref{dircos}. 
 \begin{figure}
\begin{center}
\includegraphics[height=14 cm, width=0.95\textwidth]{xsec_pol_para_fig1.eps}
 \end{center}
 \caption{{\bf (i) Top panel:} Differential scattering cross section, {\bf (ii) middle panel:} the degree of polarization ($P$), {\bf (iii) bottom panel:} the
 polar component of normalized polarization
 vector; vs the outgoing charged lepton energy ($E_\tau$) at the different values of incoming neutrino energy i.e. $E_{\nu_\tau}=6$ GeV, 10 GeV and 20 GeV. The numerical results for $\nu_\tau-N$ 
 are obtained with target mass correction effect at the next-to-leading order. These results are obtained for $W\ge1.4$ GeV and $W\ge2$ GeV cut on the center of mass energy. }
 \label{res1}
\end{figure}
\subsection{Free nucleon}
\subsubsection{Neutrino induced reactions}
In Fig.~\ref{res1}, we present the numerical results for various observables in the case of CC $\nu_\tau-$nucleon DIS process at different values of the incoming neutrino energies viz. $E_{\nu_\tau}=6,$ 10 and 20 GeV. In this figure,
the results for the double differential scattering cross section $\frac{d^2\sigma_N}{dE_\tau\;d\cos\theta}$ (top panel), degree of polarization $P$ (middle panel) and the polar component of normalized polarization
 vector $\cos\theta_P$ (bottom panel) vs. the final state charged lepton energy ($E_\tau$) are presented at different values of lepton scattering
 angle $\theta$ in the laboratory frame viz. $\theta=0^\degree,~2.5^\degree,~5^\degree$ and $10^\degree$. The results presented in this figure describe the angular and energy dependence of the above observables and the impact of applying 
 different kinematic cuts on the CoM energy $W$. In the following, we present our results for each of the above observables in case of neutrino reactions.
\begin{enumerate}[I]
 \item {\bf \underline{Differential cross section:}}\\
{\bf (i)} With the increase in neutrino energy, the deep inelastic cross section 
spans over a wide lepton energy range but has a smaller magnitude at a given lepton energy $E_\tau$ as compared to the results obtained at the lower neutrino energies. This result is applicable
for all angles in the range of $0^\degree\le\theta\le 10^\degree$ studied in this work but the reduction is smaller with the increase in $\theta$. For example, at a fixed $\tau-$lepton energy of $E_\tau=4$ GeV and $\theta=2.5^\degree$, the reduction
in the cross section at $E_{\nu_\tau}=10$ GeV is around 28\% as compared to the cross section at $E_{\nu_\tau}=6$ GeV. This reduction is observed to be $15\%$ at $\theta=10^\degree$. 
 {\bf (ii)} For a fixed neutrino energy, the $\tau-$lepton production cross section increases with $\tau-$lepton energy $E_\tau$, for a given scattering angle. 
 For example, at $E_{\nu_\tau}=10$ GeV and scattering angle $\theta=5^\degree$, the cross section increases with the $\tau-$lepton energy by $70\%$ when we move from $E_\tau=4$ GeV to $E_\tau=6$ GeV. 
 {\bf (iii)} As far as the angular dependence is concerned, for a fixed neutrino energy and for a fixed $\tau-$lepton energy, the cross section decreases as we 
 increase the scattering angle. For example, at $E_{\nu_\tau}=10$ GeV 
 and $E_\tau=4$ GeV, there is a reduction in the cross section of about $8\%$ as we move from $\theta=0^\degree$ to $\theta=5^\degree$.
{\bf (iv)} Furthermore, it may also be observed from the figure that
 due to the kinematical constraint on the CoM energy, at $E_{\nu_\tau}=10$ GeV, the $\tau-$lepton production cross section with $W\ge2$ GeV (solid line)
 gets kinematically restricted in $\tau-$lepton energy as compared to the cross section with $W\ge 1.4$ GeV (dotted line). For example, at
 $\theta= 5^\degree$, the kinematically allowed region for $\tau-$lepton energy is $2.3\le E_\tau\le8.8$ GeV for $W\ge 1.4$ GeV, which reduces to $2.6\le E_\tau \le 7.4$ GeV when a cut of $W\ge 2$ GeV is applied.
 
 \item {\bf  \underline{Degree of polarization ($P$):}}\\
 {\bf (i)} The degree of polarization ($P$) of the final
 state charged lepton (shown in the middle panel), deviates from unity for {\bf (a)} the 
 lower neutrino energies, {\bf (b)} the higher energy values of the produced $\tau-$lepton, and {\bf (c)} small scattering angles, implying that the produced $\tau-$lepton is partially polarized 
 in this kinematic region. {\bf (ii)} For a given lepton energy $E_\tau$, the degree of polarization increases with the neutrino energy $E_{\nu_\tau}$. 
 Quantitatively, for $E_{\nu_\tau}=10$ GeV and $\theta=0^\degree (5^\degree)$, the deviation from unity
 is about 3\%(4\%) at $E_\tau=4$ GeV, however, for $E_{\nu_\tau}=20 $ GeV the outgoing lepton is almost polarized with a deviation of 1-3\% for all the scattering angles considered here.
 {\bf (iii)} For a given neutrino energy $E_{\nu_\tau}$, the degree of polarization decreases with the increase in $\tau-$lepton energy, say, at $E_{\nu_\tau}=6$ GeV and $\theta=0^\degree$, the 
 deviation of $P$ from unity is 7\% for $E_\tau=3$ GeV and $25\%$ for $E_\tau=4$ GeV. 
 {\bf (iv)} For a given neutrino energy $E_{\nu_\tau}$ and lepton energy $E_\tau$, the degree of polarization increases with the lab scattering angle $\theta$. 
 For example, at $E_{\nu_\tau}=6$ GeV and $E_\tau=3$ GeV, the deviation of the degree of polarization from unity is found to be $7\%$ for the forward scattering angle $\theta=0^\degree$,
however, for $\theta=5^\degree$, it decreases to 6\% and becomes 4\% for $\theta=10^\degree$. 
  
 \begin{figure}
\begin{center}
\includegraphics[height=14 cm, width=0.95\textwidth]{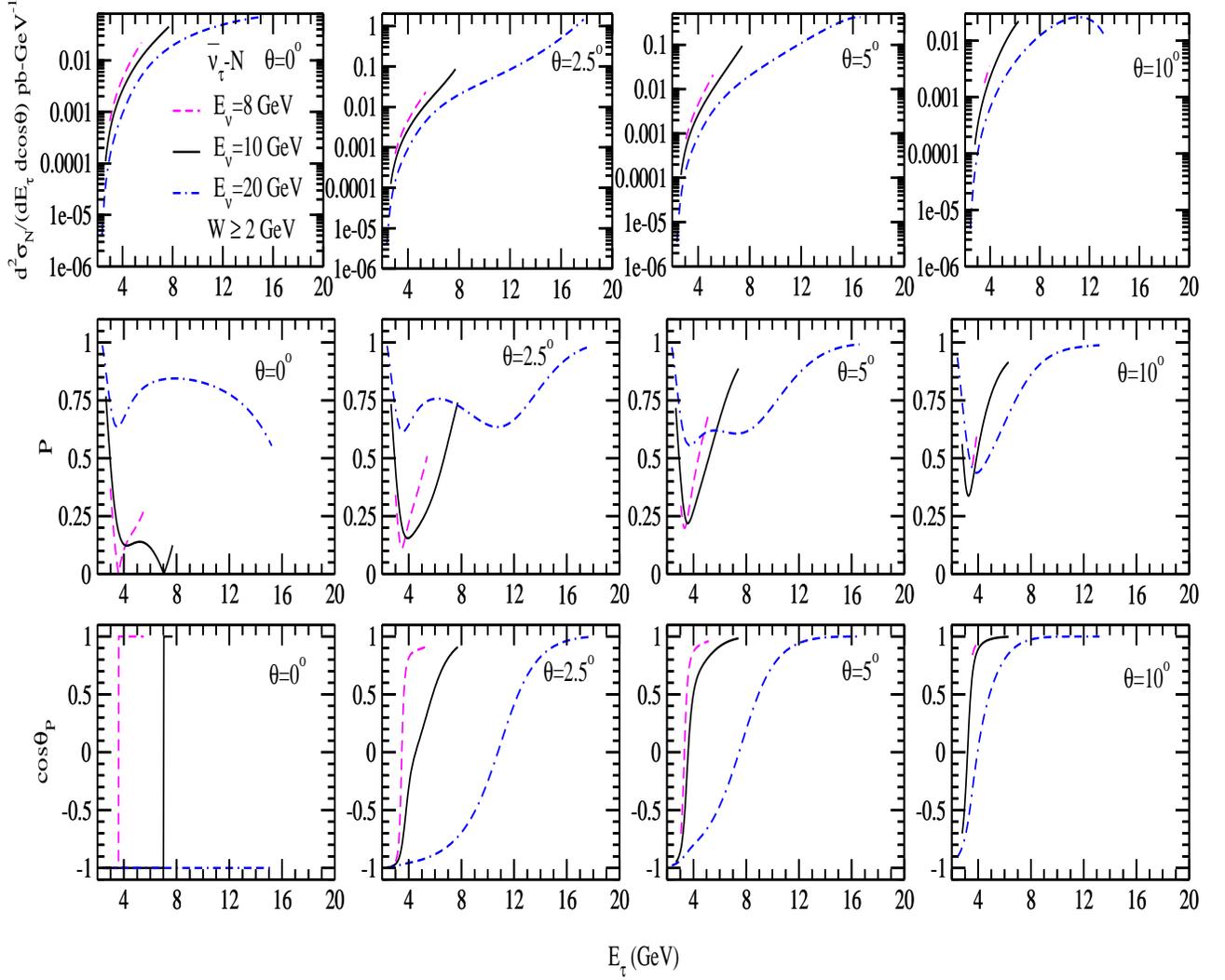}
 \end{center}
 \caption{{\bf (i) Top panel:} Differential scattering cross section, {\bf (ii) middle panel:} the degree of polarization ($P$), {\bf (iii) bottom panel:} the polar component of normalized polarization
 vector; vs the outgoing charged lepton energy ($E_\tau$) at the different values of incoming neutrino energy i.e. $E_{\nu_\tau}=8$ GeV, 10 GeV and 20 GeV. The numerical results for $\bar\nu_\tau-N$ 
 are obtained with target mass correction effect at the next-to-leading order. These results are obtained for $W\ge2$ GeV cuts on the center of mass energy. }
 \label{res2}
\end{figure}
 \item{\bf  \underline{Polar component of normalized polarization vector ($\cos\theta_P$):}}\\
  {\bf (i)} The value of the polar component of the normalized 
 polarization vector is $\cos\theta_P=+1$ at the forward scattering angle for the entire 
 range of $E_\tau$. At the non-zero values of the lepton scattering angle, $\theta$ (considered here), $\cos\theta_P$ is negative in the higher energy range of the charged lepton ($E_\tau$) while 
 for the lower energy range, the polarization vector changes its direction and $\cos\theta_P$ possesses positive values. {\bf (ii)} For a given value of $E_\tau$ and $\theta$, $\cos\theta_P$ 
 increases with the increase in neutrino energy, quantitatively,
 at $E_\tau=4$ GeV and $\theta=5^\degree$ it is found to be 15\% higher for $E_\nu=10$ GeV as compared to the case of $E_\nu=6$ GeV as shown in the figure.
\end{enumerate}
 We have found that the qualitative behavior of these results is 
 similar to the results reported by Hagiwara et al.~\cite{Hagiwara:2003di, Hagiwara:2004gs}, Aoki et al.~\cite{Aoki:2005wb} and Graczyk~\cite{Graczyk:2004vg}. However, quantitatively, there are
 some differences due to the use of different PDFs, $Q^2$ evolution, and the cuts of CoM energy $W$.

\subsubsection{Antineutrino induced reactions}
In Fig.~\ref{res2}, the corresponding results are shown for the $\bar\nu_\tau-N$ induced $\tau^+-$lepton production processes, at the different 
values of $E_{\nu_\tau}$ viz. 8 GeV (dashed line), 10 GeV (solid line) and 20 GeV (dashed-dotted line) for the kinematical conditions similar to the case 
of charged current $\nu_\tau-N$ DIS process (Fig.~\ref{res1}). These results are obtained by applying 
a cut of 2 GeV on $W$. These results for the antineutrino-nucleon interaction are discussed below.
\begin{enumerate}[I]
\item {\bf \underline{Differential cross section:}}\\
 The qualitative behavior of the scattering cross section is similar for both $\nu_\tau-N$ and $\bar\nu_\tau-N$ DIS processes while quantitatively the $\tau^+-$lepton production 
cross sections are smaller as compared to the case of $\tau^--$lepton production. For example, at the fixed values of $E_{\nu_\tau}=10$ GeV and $E_\tau=6$ GeV, this suppression 
is found to be about 82\% at $\theta=0^\degree$, 80\% at $\theta=2.5^\degree$, and 70\% at $\theta=10^\degree$.

 \item {\bf \underline{Degree of polarization ($P$):}}\\
 {\bf (i)} The results obtained for the degree of polarization for the $\tau^+-$lepton are 
different from the corresponding results for the $\tau^--$lepton polarization. It may be noticed from the figure that at the lower values of $E_\tau$, the outgoing $\tau^+-$lepton 
possesses a good strength of the degree of polarization $P$. As we move towards the higher values of $E_\tau$, the value of $P$ gets reduced until it approaches the minimum at which
$\tau^+$ is likely to be less polarized. However, with the further increase in $E_\tau$, the degree of polarization increases again. 
We find that at the forward scattering angle, i.e., 
$\theta=0^\degree$, and at $E_{\nu_\tau}=10$ GeV, the produced $\tau^+-$lepton 
possesses about $46\%$ of the degree of polarization at $E_\tau=3$ GeV which reduces to 11\% at $E_\tau=6$ GeV and at $E_\tau=7$ GeV, $\tau^+$ becomes almost unpolarized as the value of $P$ 
is negligible here. 
{\bf (ii)} We also observe that with the increase in the neutrino beam energy, the degree of polarization 
increases, e.g., at $E_\tau=6$ GeV and at $\theta=2.5^\degree$, it is found to be
36\% for $E_{\nu_\tau}=10$ GeV and 75\% for $E_{\nu_\tau}=20$ GeV.
{\bf (iii)} For a given value of neutrino energy and the $\tau-$lepton energy, the degree of polarization increases with the increase in the lepton scattering angles. Quantitatively,
for $E_{\nu_\tau}=10$ GeV and $E_\tau=6$ GeV, the degree of polarization of the $\tau^+-$lepton is 36\% at $\theta=2.5^\degree$, 67\% at $\theta=5^\degree$ and 90\% at $\theta=10^\degree$.

 \item {\bf \underline{Polar component of normalized polarization vector ($\cos\theta_P$):}}\\
 {\bf(i)} A comparison with Fig.~\ref{res1} shows that the direction of polarization vector for $\tau^+-$lepton is opposite to the direction of 
$\tau^--$lepton. {\bf(ii)} The polarization vector of the produced $\tau^+-$lepton has negative values of $\cos\theta_P$ (implying $\tau-$lepton to be almost left handed)
when the charged lepton energy is low, while for the high energy values, the direction of the polarization vector
gets flipped ($\cos\theta_P \ge 0$). {\bf(iii)} For a given neutrino energy $E_\nu$ and charged lepton energy $E_\tau$, the value of $\cos\theta_P$ increases with the increase in the scattering angle.
Quantitatively, at $E_\nu=10$ GeV and $E_\tau=5$ GeV, there is an increase of about $77\%$ in the value of $\cos\theta_P$  at $\theta=5^\degree$ and $\sim 81\%$ at $\theta=10^\degree$ as compared to 
the case of $\theta=2.5^\degree$. 
\end{enumerate}
 We observe that qualitatively our results for $\bar\nu_\tau$ induced processes also show similar behavior 
as the corresponding results reported by Hagiwara et al.~\cite{Hagiwara:2003di, Hagiwara:2004gs},
although the quantitative difference is more pronounced in this case as compared to the case of $\nu_\tau-N$ DIS discussed earlier in the text.
\begin{figure}[h]
\begin{center}
\includegraphics[height=8 cm, width=14 cm]{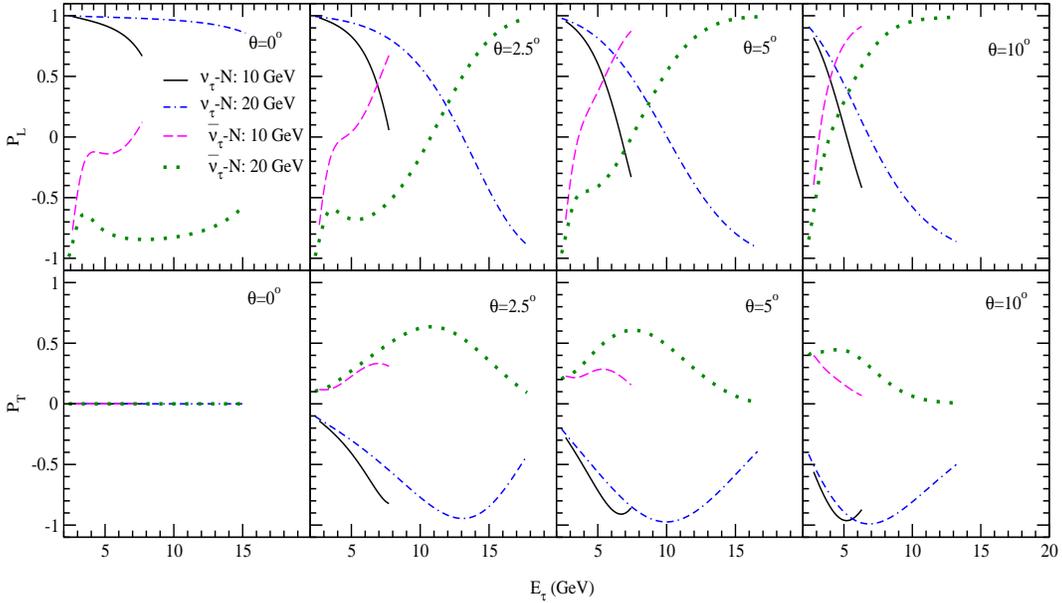}
 \end{center}
\caption{{\bf (i) Top panel:} Longitudinal component of the polarization vector ($P_L$), {\bf (iii) bottom panel:} transverse component of the polarization vector ($P_T$); vs the outgoing charged
 lepton energy ($E_\tau$) at $E_{\nu_\tau}=10$ GeV and $E_{\nu_\tau}=20$ GeV for $\nu_\tau/\bar\nu_\tau-N$. The numerical results are obtained with target mass correction effect at the next-to-leading order
 by considering $W\ge 2$ GeV.}
 \label{res3}
\end{figure}
\subsubsection{Polarization components $P_{L,T}(\tau^-)$ and $P_{L,T}(\tau^+)$}
We explicitly present the results for $P_L$ (top panel) and $P_T$ (bottom panel) vs $E_\tau$ in Fig.~\ref{res3} for various angles $\theta$ in the range 
of $0^\degree\le \theta \le 10^\degree$.
These results are presented for both the $\tau^-$ and $\tau^+$ leptons produced respectively in the 
$\nu_\tau$ and $\bar\nu_\tau$ induced DIS processes at $E_{\nu_\tau}=10$ GeV and $E_{\nu_\tau}=20$ GeV in order to show the energy dependence. 
\begin{enumerate}[I]
\item {\bf \underline{Longitudinal polarization component $P_L$} }\\
{\bf (i)} For a given lepton energy $E_\tau$ and $\theta$, the value of $P_L(\tau^\mp)$ increases with the neutrino energy $E_{\nu_\tau}$. {\bf(ii)} 
For a given neutrino energy the polarization component $P_L(\tau^-)$ is large and positive (almost +1) at the smaller $\tau^--$lepton energy $E_\tau$
and it becomes negative and large (approaching -1) with the increase in $E_\tau$. This behavior of $P_L(\tau^-)$ to approach the value of -1 at the higher lepton energies is faster at larger angles.
{\bf (iii)} In the case of $\tau^+-$lepton production, the behavior of $P_L(\tau^+)$ is qualitatively similar to the behavior of $\tau^--$lepton production except that the sign of $P_L(\tau^+)$
for $\tau^+$ is opposite of the sign of $P_L(\tau^-)$ for $\tau^-$. Quantitatively, the nature of the energy dependence with $E_{\nu_\tau}$ and $E_\tau$, and $\theta-$dependence of $P_L(\tau^-)$ and $P_L(\tau^+)$
are different specially at lower scattering angles $\theta$ as shown in the figure.
\begin{figure}[h]
\begin{center}
\includegraphics[height=12 cm, width=0.95\textwidth]{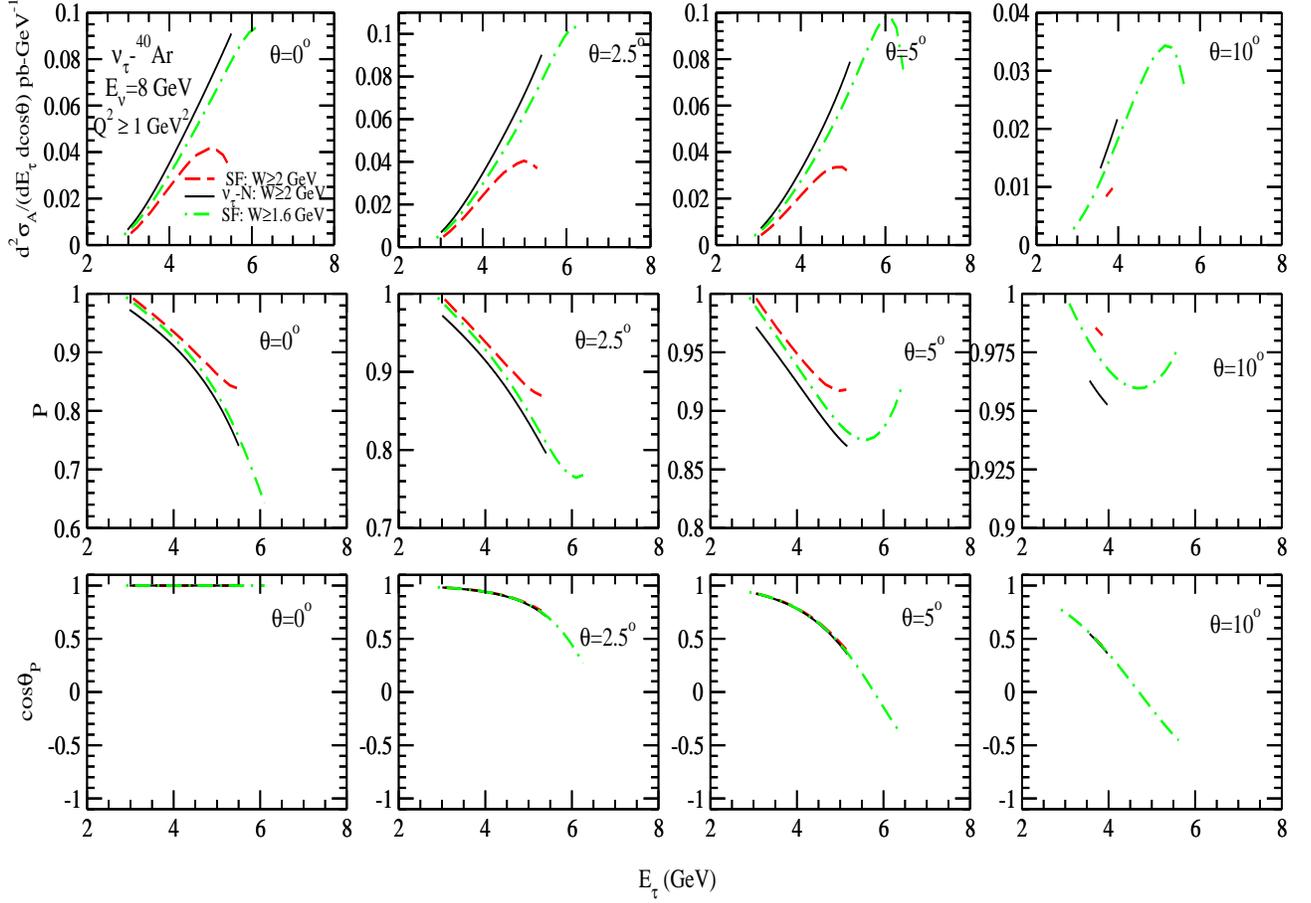}
 \end{center}
\caption{{\bf (i) Top panel:} Differential scattering cross section, {\bf (ii) middle panel:} the degree of polarization ($P$), {\bf (iii) bottom panel:} the polar component of normalized polarization
 vector vs the outgoing charged lepton energy ($E_\tau$) at $E_{\nu_\tau}=8$ GeV. The numerical results for $\nu_\tau-^{40}$Ar are obtained using spectral function only (labeled as 'SF') 
 with target mass correction effect at the next-to-leading order by applying a cut of $W\ge 1.6$ GeV (dashed-dotted line) and $W\ge 2$ GeV (long dashed line). These results are compared
 with the results of free nucleon case (solid line) obtained with a cut of 2 GeV on the CoM energy $W$.}
 \label{res4}
\end{figure}
\item {\bf \underline{Transverse polarization component $P_T$} }\\
In case of the transverse polarization of $\tau-$lepton, {\bf (i)} $P_T$ vanishes at $\theta=0^\degree$. With the increase in scattering angle $\theta$, $P_T$ for $\tau^-(\tau^+)$ increases 
with energy $E_\tau$ taking negative (positive) values of $P_T$. This increase of $P_T$ with energy $E_{\tau^-(\tau^+)}$ continues until a critical energy $E_{\tau^-(\tau^+)}^{C}$, after 
which it decreases. The value of $E_{\tau^-}^C$ and $E_{\tau^+}^C$ depends upon the incident neutrino energy $E_{\nu_\tau}$ of $\nu_\tau$ and $\bar\nu_\tau$. In general, $E_{\tau^-}^C$ is different than
$E_{\tau^+}^C$ for a given energy $E_{\nu_\tau}$ for neutrinos and antineutrinos as shown in Fig.~\ref{res3} (lower panel).
{\bf (ii)} The angular dependence of $P_T$ on $\theta$ for nonzero values of $\theta$ first increases and then decreases with the increase in angle $\theta$ in the region of 
$2.5^\degree\le \theta \le 10^\degree$ as shown in the figure. {\bf (iii)} The non-vanishing values of $P_T$ at $\theta\ne 0^\degree$ decrease with the increase in energy $E_{\nu_\tau}$, for example,
the decrease in $P_T$ is about $20\%$ for $E_{\nu_\tau}=20$ GeV as compared to $E_{\nu_\tau}=10$ GeV, at $\theta=2.5^\degree$ and $E_\tau=6$ GeV.

The numerical values of $P_L$ and $P_T$ and their lepton energy as well as angular dependences are qualitatively similar to the energies and angular dependence of $P_L$ and $P_T$ seen in the case of 
quasielastic reactions~\cite{Kuzmin:2004yb}.
\end{enumerate}

\begin{figure}
\begin{center}
\includegraphics[height=12 cm, width=0.95\textwidth]{xsec_pol_para_nutau_10gev_argon_sf.eps}
 \end{center}
 \caption{{\bf (i) Top panel:} Differential scattering cross section, {\bf (ii) middle panel:} the degree of polarization ($P$), {\bf (iii) bottom panel:} the polar component of normalized polarization
 vector vs the outgoing charged lepton energy ($E_\tau$) at $E_{\nu_\tau}=10$ GeV for the charged current $\nu_\tau-^{40}Ar$ DIS process. Lines in this figure have the same meaning as in Fig.~\ref{res4}.}
 \label{res5}
\end{figure}

\subsection{Nuclear effects in $^{40}Ar$}
\begin{figure}
\begin{center}
\includegraphics[height=12 cm, width=0.95\textwidth]{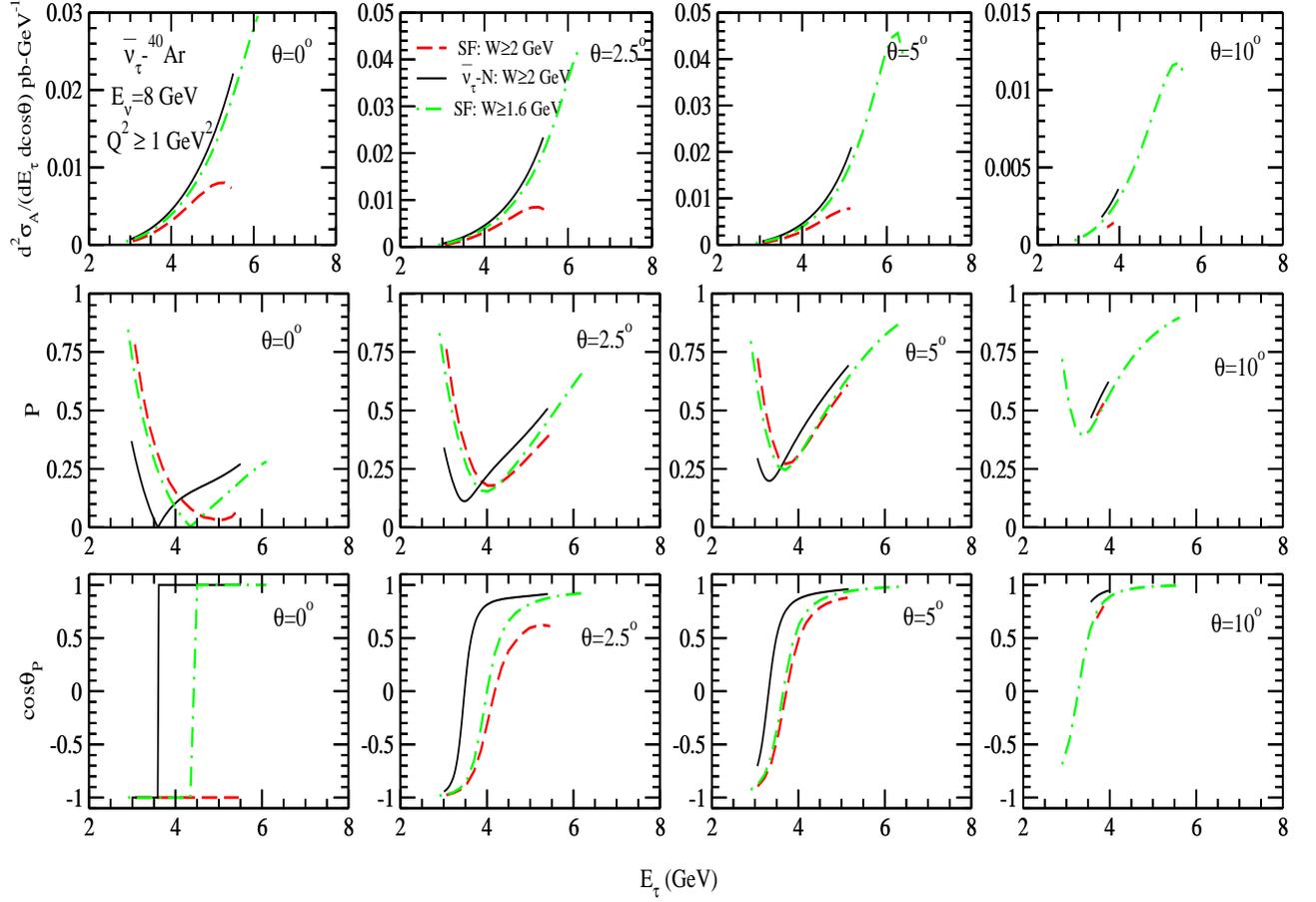}
 \end{center}
\caption{{\bf (i) Top panel:} Differential scattering cross section, {\bf (ii) middle panel:} the degree of polarization ($P$), {\bf (iii) bottom panel:} the polar component of normalized polarization
 vector vs the outgoing charged lepton energy ($E_\tau$) at $E_{\nu_\tau}=8$ GeV for the charged current $\bar\nu_\tau-^{40}Ar$ DIS process. Lines in this figure have the same meaning as in Fig.~\ref{res4}.}
 \label{res6}
\end{figure}

We have taken into account the nuclear medium effects such as the Fermi motion, the Pauli blocking and the nucleon correlations through the 
nucleon spectral function which is properly normalized to the mass number of $^{40}Ar$ (using Eq.~\ref{shnorm}). The obtained numerical results are presented in Figs.~\ref{res4}-\ref{res7}
for $\nu_\tau$ and $\bar\nu_\tau$ induced reactions in $^{40}Ar$.
\subsubsection{Neutrino induced reactions}\label{nusub}

In Figs.~\ref{res4} and \ref{res5}, the results are presented for the $\tau^--$lepton production in the charged current $\nu_\tau-^{40}Ar$ DIS process for the projectile beam energies of 
$E_{\nu_\tau}=8$ GeV and $E_{\nu_\tau}=10$ GeV, respectively. We observe that:

{\bf [I] \underline{Differential cross section:}}\\
 {\bf (i)} The effects of the nuclear medium modifications are larger in the
higher region of the $\tau^--$lepton
energy and specially at the larger scattering angles. For example, when the $\tau-$lepton energy is $E_\tau=7$ GeV and $E_{\nu_\tau}=10$ GeV, the differential cross section gets suppressed 
due to the nuclear medium effects which is approximately by 
$24\%$ at $\theta=2.5^\degree$, and $36\%$ at $\theta=5^\degree$.
\begin{figure}[h]
\begin{center}
\includegraphics[height=12 cm, width=0.95\textwidth]{xsec_pol_para_nutaubar_10gev_argon_sf.eps}
 \end{center}
\caption{{\bf (i) Top panel:} Differential scattering cross section, {\bf (ii) middle panel:} the degree of polarization ($P$), {\bf (iii) bottom panel:} the polar component of normalized polarization
 vector vs the outgoing charged lepton energy ($E_\tau$) at $E_{\nu_\tau}=10$ GeV for the charged current $\bar\nu_\tau-^{40}Ar$ DIS process. Lines in this figure have the same meaning as in Fig.~\ref{res4}.}
 \label{res7}
\end{figure}
 {\bf (ii)} The nuclear medium effects become smaller with the increase in neutrino energy and larger with the increase in scattering angle $\theta$.
 For example, when the neutrino beam energy is $E_{\nu_\tau}=8$ GeV, 
the difference in the results of the differential cross sections obtained for the free nucleon target and the argon nuclear target, due to the nuclear medium
modifications at $E_\tau=3$ GeV, is found to be 28\%(40\%) which becomes 15\%(22\%) for $E_{\nu_\tau}=10$ GeV
when the lepton scattering angle is $\theta=0^\degree(5^\degree)$. 
 {\bf (iii)} The allowed kinematic region for $E_\tau$ becomes more restrictive for higher cut of CoM energy $W$. Quantitatively, when $W\ge 1.6$ GeV cut is applied on the differential 
 cross section obtained for $E_{\nu_\tau}=8$ GeV at $\theta=2.5^\degree$,
 the allowed kinematic region of the $\tau-$lepton energy is $2.9\le E_\tau\le6.2$ GeV which reduces to $3 \le E_\tau\le 5.3$ GeV for a cut of $W\ge 2$ GeV. 
 {\bf (iv)} For a given $\tau-$lepton energy $E_\tau$, the differential cross section for $\nu_\tau-^{40}Ar$ scattering process reduces due to the nuclear medium effects at higher values of
 cut on the CoM energy $W$ unlike the scattering of $\nu_\tau$ from a free nucleon target.
For example, the results of the differential cross section obtained for $^{40}Ar$ nuclear target at $E_\tau=3$ GeV, with a cut of 
$W\ge 2$ GeV, are approximately 32\% suppressed as compared to the cross section obtained with $W \ge 1.6$ GeV cut when the neutrino beam energy is $E_\nu=8$ GeV and $\theta=5^\degree$. 

{\bf [II] \underline{Degree of polarization (P):}}\\
{\bf(i)} There is very little change in the degree of polarization due to the nuclear medium effects at lower $\tau-$lepton energies. For example, at $E_\tau \le 6$ GeV, there is an increase 
of about $1-2\%$ in the results of argon as compared to the free nucleon results for all the scattering angles taken into consideration at $E_{\nu_\tau}=10$ GeV. 
However, for the higher values of $\tau-$lepton energy, say $E_\tau=7.5$ GeV it is found to be 9\% at $\theta=0^\degree$ and 5\% at $\theta=2.5^\degree$. 
{\bf(ii)} We observe that the effect of the nuclear medium on the degree of polarization decreases for the larger scattering angles and higher neutrino energies.


{\bf [III] \underline{Direction of polarization vector:}}\\
It is interesting to note that the effects of the nuclear medium on the direction of polarization vector are very small. 
The direction of the polarization vector is almost independent of the choice of 
kinematic constraint on the center of mass energy. 

\subsubsection{Antineutrino induced reactions}
In Figs.~\ref{res6} and \ref{res7}, the results are presented for $\bar\nu_\tau-^{40}Ar$ DIS process.
In the case of $\tau^+-$lepton production induced by $\bar\nu_\tau$ the nuclear medium effects are found qualitatively similar to the $\tau^--$lepton production induced by $\nu_\tau$.
However, quantitatively there are some differences, for example: 

{\bf [I] \underline{Differential cross section:}}\\
{\bf(i)} We observe that for the projectile beam energy of $E_{\nu_\tau}=8$ GeV, due to the nuclear medium effects the differential cross section for the $\tau^+-$lepton 
production are reduced by about 34\% at $\theta=2.5^\degree$ and 45\% at $\theta=5^\degree$, when $E_\tau$ is fixed at 3 GeV. {\bf(ii)} At the higher values of $E_{\tau}$, the nuclear medium effects become more 
pronounced, for example, at $E_\tau=5$ GeV, the reduction in the differential cross section for argon is found to be 47\% at $\theta=2.5^\degree$ and 55\% at $\theta=5^\degree$, as compared to the
case of free nucleon.
{\bf (iii)} With the increase in $E_{\nu_\tau}$, the nuclear medium effects get reduced similar to the case of $\tau^--$lepton production. Quantitatively, 
for $E_{\nu_\tau}=10$ GeV there is a reduction of about 30\% at $\theta=0^\degree$ and $41\%$ at $\theta=5^\degree$ as compared to the case of $E_{\nu_\tau}=8$ GeV when $E_\tau$ is fixed at 5 GeV.
{\bf (iv)} The nuclear medium effects for $\tau^+-$lepton production are found to be quantitatively different than the case of $\tau^--$lepton production. For example, there is a difference 
of about 4-5\% at $E_\tau=3$ GeV and approximately 3-4\% at $E_\tau=5$ GeV in the range of $0^\degree\le \theta \le 10^\degree$.
{\bf (v)} The impact of the CoM energy cut on the $\tau^+-$lepton production cross section is found to be qualitatively similar to the case of $\tau^--$lepton production
as stated in subsection B.1 [I](iv).

{\bf [II] \underline{Degree of polarization (P):}}\\
{\bf(i)} The $\tau^+-$lepton has a comparatively lower degree of polarization at the moderate values 
of $E_\tau$, however, at the extreme ends, it has a good strength of $P$ except at $\theta=0^\degree$. {\bf(ii)} In the low and intermediate energy region of $\tau^+-$lepton, the degree of 
polarization increases due to the nuclear medium effects, while in the higher region of $E_\tau$ it decreases for $\theta\ne0$. For example,
at $E_{\nu_\tau}=8$ GeV and $\theta=2.5^\degree$,
the nuclear medium effects on the degree of polarization are approximately $55\%$ for $E_\tau=3$ GeV and $37\%$ for $E_\tau=5$ GeV.

{\bf [III] \underline{Direction of polarization vector:}}\\
{\bf(i)} The value of $\cos\theta_P$ for the $\tau^+-$lepton decreases due to the nuclear medium effects except at $\theta=0^\degree$.
For example, on comparing the results obtained with a cut of $W\ge 2$ GeV at $\theta=0^\degree$, we observe that the value of 
$\cos\theta_P$ remains -1 in the entire range of $E_\tau$ when $\bar\nu_\tau$ interacts with a nucleon bound inside the argon nuclear target, however, in the case of free nucleon
target $\cos\theta_P$ is -1 up to a certain value of 
$E_\tau$ after which the direction of polarization vector gets flipped and it becomes +1. {\bf(ii)} It is also noticeable that at $\theta=0^\degree$, the results obtained for 
$^{40}Ar$ with a lower cut on $W$, i.e., 1.6 GeV also shows the similar behavior as observed in the case of free nucleon target. {\bf(iii)} The nuclear medium effects 
on $\cos\theta_P$ for $\tau^+-$lepton production
are found to be larger than observed in the case of $\tau^--$lepton production. Quantitatively, for $E_\nu=10$ GeV and $\theta=2.5^\degree$, the nuclear medium effects are approximately $60\%$ larger for 
$\tau^+-$lepton as compared to $\tau^--$lepton at $E_\tau=4$ GeV
and $\sim 20\%$ larger at $E_\tau=6$ GeV.

To conclude, in this work we have studied $\tau^\pm-$lepton polarization in the presence of nuclear medium effects for the charged current  
$\nu_\tau/\bar\nu_\tau$-$^{40}Ar$ deep inelastic scattering. This study may be helpful in reducing the existing systematic uncertainties in the determination of the (anti)neutrino-nucleon and 
(anti)neutrino-nucleus scattering cross sections. Moreover, this work may be useful for ongoing and upcoming experiments such as FASER$\nu$, SND$@$LHC, DUNE, and IceCube upgrade in
estimating the background events in the $\nu_\mu \to \nu_e$ appearance mode as well as in performing the more precise measurements of the scattering cross sections. We 
plan to study the decay distributions of the polarized charged $\tau$-lepton and will report them in future communications.

\section*{Acknowledgment}   
F. Zaidi is thankful to Council of Scientific \& Industrial Research, Govt. of India for providing Senior Research Associateship (SRA) under the Scientist's Pool Scheme, file no. 13(9240-A)2023-POOL and to 
the Department of Physics, Aligarh Muslim University, Aligarh for providing the necessary facilities to pursue this research work.
M. S. A. is thankful to the Department of Science and Technology (DST), Government of India for providing 
financial assistance under Grant No. SR/MF/PS-01/2016-AMU/G.



%
%
%
%
%
%
%
%
%
%
%
%
%
\end{document}